\documentclass[preprint]{aastex}

\begin{document}
\title{Chemical Abundance Inhomogeneities in Globular Cluster Stars}
\author{Judith G. Cohen\altaffilmark{1}}

\altaffiltext{1}{Palomar Observatory, Mail Stop 105-24,
California Institute of Technology, Pasadena, Ca., 91125, 
jlc@astro.caltech.edu}

\begin{abstract}
It is now clear that abundance variations from star-to-star among
the light elements, particularly C, N, O, Na and Al, are ubiquitous
within galactic globular clusters; they appear seen whenever data of 
high quality is obtained for a sufficiently large sample of stars 
within such a cluster.  The correlations and anti-correlations among 
these elements and the range of variation of each element appear to be 
independent of stellar evolutionary state, with the exception that 
enhanced depletion of C and of O is sometimes seen just at the RGB tip.
While the latter behavior is almost certainly due to internal
production and mixing,  the internal mixing hypothesis can now be ruled out
for producing the bulk of the variations seen.  We focus on the implications 
of our new data for any explanation invoking primordial variations in the 
proto-cluster or accretion of polluted material from a neighboring AGB star.
 
\end{abstract}

%

Over the past two decades the
upper giant branches of the nearer GCs have been well studied
with 4-m class telescopes 
by, among others, the Lick-Texas group (see Sneden et al 2004
and references therein) or the Padua group (Gratton and his collaborators,
see, e.g. Carretta \& Gratton 1997)
(see also the early review of some of the issues to be
discussed here by Kraft 1994 and the recent review
of Gratton, Sneden \& Carretta 2004).  Recently
10-m class telescopes coupled with 
efficient spectrographs have enabled us to explore 
detailed abundance ratios and chemical history
ever deeper in the stellar luminosity function in galactic globular
clusters (GCs).  We can now reach
with considerable precision the RGB in {\it{all}} galactic GCs
(see, for example, the study of NGC~7492, 
at a distance of 26 kpc, by Cohen \& Melendez 2005b).
For the nearer GCs, abundance analyses for the  brightest main sequence stars
in the nearest GCs,
and for the subgiant branch for those slightly more distant, are now feasible.

The chemical analyses within the past 5 years
in which the author has been involved include the GCs
NGC 6528 (Carretta et al 2001), NGC 6533, M71, 
M5 (Ram\'{\i}rez \& Cohen 2003), M3, M13
(Cohen \& Melendez 2005a and reference therein) and shortly M15 and M92
as well as NGC~7492 (Cohen \& Melendez 2005b) and Pal~12 (Cohen 2004), 
all observed with HIRES (Vogt et al. 1994)
at Keck.  Our approach is to study stars over the full
range of luminosity from the RGB tip to the faintest possible that
can be reached in 2 to 4 hours of integration.  
The large program carried out with UVES at the VLT 
described in Gratton et al (2001) concentrate
on comparing subgiants with a small number of main sequence stars
in the nearest southern clusters.

Viewing in totality the collective effort of these and other groups,
we find that star-to-star abundance variations from star-to-star among
the light elements, particularly C, N, O, Na and Al, are ubiquitous.
They are
seen whenever data of sufficient quality is obtained for a sufficiently
large sample of stars within a galactic globular cluster.
Our most recent Na-O anti-correlation, for M13,
with a sample of 25 stars reaching almost to the main sequence turnoff
can be seen in Fig.~16 of Cohen \& Melendez (2005a), not shown
here due to lack of space.

The correlations and anti-correlations among these elements
for stars in GCs
and the range of variation of each element 
resemble those of proton-burning.  They
appear to be independent
of stellar evolutionary state, with the exception that enhanced
depletion of C and of O is sometimes seen just at the RGB tip.
This extra depletion of O just near the RGB tip is seen in
our M13 data shown (see also Sneden et al 2004).
Metal poor halo field stars, however, show no evidence for
O burning (Gratton et al 2000) or Na enhancement.  The variations
seen in the field stars are much closer to those predicted by
classical stellar evolution that those seen in the GC stars.

At the same time, the abundance ratios among the elements heavier
than Al, at least through the Fe peak, do not show any detectable
variation in any known GC (except, of course, $\omega$~Cen).  The
rock steady abundances for these elements requires explanation
as well, and places  important constraints on the formation mechanisms
of GCs.

Interesting as this is, we are still plagued, when observing
at high dispersion, with small samples, at least until
FLAMES came into use.  Small samples trying to discern small
variations is not the ideal combination.

My approach to this issue has been to use the molecular bands of
CH, CN and now NH to study the star-to-star abundance variations
of C and of N.  Since these bands are strong enough to be observed
at moderate resolution, I can use the multiplexing capability
of the Low Resolution Imaging Spectrograph at Keck (Oke et al 1995)
to build up large samples.  This effort is being undertaken jointly
with Michael Briley of the University of Wisconsin at Oshkosh and
with Peter Stetson of the National Research Council, Victoria, Canada.

We have now analyzed large samples of low luminosity stars (subgaints
or main sequence turnoff region stars) in each of
four GCs spanning a wide range in metallicity.  In each cluster we
have a sample of $\sim$70 stars.   Our most recent
work, a study of M15, is being written up for publication.
Our analyses of M71, M5 and M13 are already published (see
Cohen, Briley \& Stetson 2002, Briley, Cohen \& Stetson
2002 and 2004b, and references therein).  We
derive C/Fe from the CH band, and N/Fe from the CN band.  For 
M15 we must use the NH band at 3360~\AA; the CN bands are too weak.  This has
the advantage of achieving a N/H ratio which is to first
order independent of the C abundance, which would not hold were CN to be used
for this purpose.

\begin{figure}
\begin{center}
\includegraphics[width=.95\textwidth]{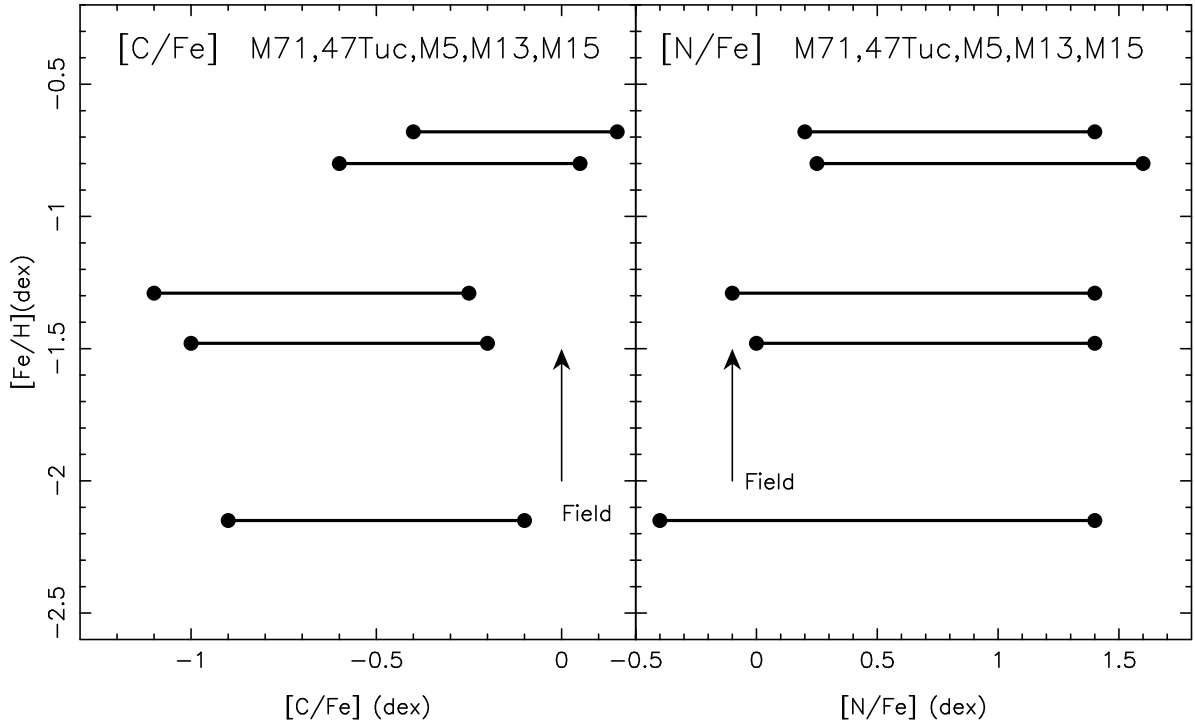}
\end{center}
\caption[]{The range of [C/Fe] (left panel) and [N/Fe] 
(right panel) is shown as a function of metallicity
([Fe/H]) for the globular clusters 
from our work on M71,  M5, M13, and M15 as well as 
for 47 Tuc (from Briley et al 2004a).  Large samples of
stars, mostly subgiants, were used in each case.
Each GC is represented by a horizontal line.
The characteristic field star ratio, from Carretta, Gratton \& Sneden (2000)
for C and from Henry, Edmunds \& Koppen (2000) for N,
are indicated by vertical arrows in each panel. }
\label{m15}
\end{figure}

Fig.~10 from our study of a large sample of stars in M5
(Cohen, Briley \& Stetson 2002) (not reproduced here due to
limits of length) is typical of the GCs studied
thus far in such detail.  It shows a strong anti-correlation
between C and N abundances, i.e. conversion of C into
N, with strong-to-star variations 
in derived C and N abundances seen at all luminosities probed.  (This
sample contains mostly stars at the base of the RGB and
just below the main sequence turnoff, with  V $\sim$ 16.5 to 19 mag, 
where the turnoff
of M5 is at about 18.2 mag.) We also find ON burning
is required to reproduce  the most extreme N enhancements, which
are very large.  In our paper, we
commented that external pollution from a nearby AGB star,
presumably a binary companion, perhaps can match the star-to-star
abundance patterns seen in GCs, but
does not seem capable of producing the highly organized
abundance variations as such an ``external'' mechanism is stochastic in nature.
Furthermore, the amount of ``polluted'' mass that needs to be accreted becomes
a significant fraction of the total mass of the
low luminosity star.  The popular
AGB companion hypothesis is beginning to crumble at this point.

Combining all our data with the study of 47 Tuc by Briley et al (2004a), we 
determine roughly the range in
[$X$/Fe], where $X$ is either C or N, in each GC.  The results
are shown in Fig.~1; the vertical arrows indicate the mean
location of metal poor field stars.  Note that the
field star mean coincides roughly with the GC high end of
the range for C and with the low end of the GC range for N.
These GCs span a range
in metallicity of a factor of $\sim$40.  Yet we find that this
range is approximately constant  for both C and for N.
Thus the additional material is not from some primary process which
dumps a fixed amount of N into the GC gas.  Instead it behaves like
a secondary process, increasing as [Fe/H] increases.  Since the
production of C and N in AGB stars is to first order a primary
process, this strongly suggests that ejecta from AGB stars do not cause
the star-to-star variation in the abundance of these elements in GCs.
This leaves some kind of variation imprinted in the proto-cluster
before the present generation of stars we now observe were formed
as the only viable scenario.  Furthermore
the source of this cannot have been some previous generation
of AGB stars, unless mass loss
rates vary proportionately to metallicity.  While it is believed
that they do increase with metallicity, the factor generally
discussed is far smaller than the factor of 40 range in our GC sample.

So after more than 20 years of searching for an answer, we have much
better data on the nature of star-to-star variations in the abundances
of the light elements in GCs, but still no definitive understanding
of the physical mechanism(s) responsible, nor of why metal poor field halo
stars do not show these phenomena.

%

\end{document}